\documentclass[aps,prd,final, nofootinbib]{revtex4}
\usepackage{amsmath, amsfonts, amsthm, amssymb, graphicx,epsfig}
\usepackage{subfig}
\usepackage{epstopdf}
\usepackage{color}
\def\be{\begin{equation}}
\def\ee{\end{equation}}
\def\ba{\begin{eqnarray}}
\def\ea{\end{eqnarray}}

\def\bdm{\begin{displaymath}}
\def\edm{\end{displaymath}}

\def\bq{\begin{quote}}
\def\eq{\end{quote}}

\def\d{{\rm d}}
\def\del{\partial}
\def\ltap{\ \raise.3ex\hbox{$<$\kern-.75em\lower1ex\hbox{$\sim$}}\ }
\def\gtap{\ \raise.3ex\hbox{$>$\kern-.75em\lower1ex\hbox{$\sim$}}\ }
\def\gl{\ \raise.5ex\hbox{$>$}\kern-.8em\lower.5ex\hbox{$<$}\ }
\def\roughly#1{\raise.3ex\hbox{$#1$\kern-.75em\lower1ex\hbox{$\sim$}}}

 at 11truept

\def\d2{{(\hat D \pi)^2 }}

\def \n {{\nabla}}

\newcommand{\beq}{\begin{equation}}
\newcommand{\eeq}{\end{equation}}
\newcommand{\bea}{\begin{eqnarray}}
\newcommand{\eea}{\end{eqnarray}}
\newcommand{\beqa}{\begin{eqnarray}}
\newcommand{\eeqa}{\end{eqnarray}}
\newcommand{\nn}{\nonumber\\}

\def \1 {\pi_{i_1}}
\def \2 {\pi_{i_2}}
\def \3 {\pi_{i_3}}
\def \4 {\pi_{i_4}}
\def \5 {\pi_{i_5}}
\def \6 {\pi_{i_6}}
\def \7 {\pi_{i_7}}
\def \m {\pi_{i_m}}
\def \a {\hat a}
\def \b {\hat b}
\def \c {\hat c}

\begin{document}

\title{Covariant multi-galileons and their generalisation}

\author{Antonio Padilla}
\email[]{antonio.padilla@nottingham.ac.uk}
\author{Vishagan Sivanesan}
\email[]{ppxvs@nottingham.ac.uk}
\affiliation{School of Physics and Astronomy ,University of Nottingham, Nottingham NG7 2RD, UK} 
\date{\today}

\begin{abstract}
We find a covariant completion of the flat-space multi-galileon theory, preserving second-order field equations. We then generalise this to arrive at  an enlarged class of second order theories describing multiple scalars and a single tensor, and conjecture that these are a multi-scalar version of Horndeski's most general scalar-tensor theory.
\end{abstract}


\maketitle

\section{Introduction}
Whilst General Relativity is supremely successful at describing gravity at solar system scales, it can only be made compatible with observation at galactic and cosmological scales if we accept that 95\% of the Universe is composed of stuff we know very little about, namely dark matter \cite{DM} and dark energy \cite{DE}.  Dark energy is particularly poorly understood from a particle physics perspective, so for this reason it is natural to ask if General Relativity ought to be replaced by a modified theory of gravity at the relevant scales (see \cite{review} for an extensive review). The simplest modifications of gravity are those that contain additional scalar fields. So called scalar-tensor theories date back to Scherrer \cite{Scherrer}, Jordan \cite{Jordan}, and Thiry \cite{Thiry}, and, most notably, to Brans and Dicke \cite{BD}.  Higher dimensional gravity generically introduces additional scalar fields when reduced down to four dimensions, through either Kaluza-Klein or braneworld compactifications (for reviews, see, \cite{Grana,roy,thesis}).

In 1974, Horndeski derived the {\it most general} theory describing a single scalar and a single tensor in four dimensions, preserving second order field equations.  This last condition is required in order to avoid Ostrogradski ghosts associated with higher derivatives \cite{ostro}. Much more recently, DGSZ  \cite{dgsz} developed a generalisation of covariant galileon theory \cite{gal,covgal}. The DGSZ model was later shown to be equivalent to Horndeski's theory in four dimensions \cite{Kob}, although it is written in a more elegant form and generalises to any number of dimensions.  Recent interest in Horndeski's theory has been considerable, ranging from applications to inflation \cite{Kob,higgs}, a discussion of the Vainshtein mechanism \cite{hornvainsh}, and the derivation of boundary terms and junction conditions \cite{hornjunc}. We should also note that  it contains a number of very interesting theories as a subset including k-essence \cite{kessence}, covariant-galileons \cite{covgal}, KGB gravity \cite{kgb} and the self-tuning {\it Fab-Four} scenario \cite{fab4}. 

In this paper we work towards a multi-scalar analogue of Horndeski's theory, describing the most general theory of multiple scalars and a single tensor, admitting second order field equations.  We begin with the multi-galileon theory described in $D$ dimensional Minkowski space\footnote{This is the unique action that (i) is invariant under so-called galilean symmetry $\pi \to \pi +b_a x^a +c$ and (ii) has at most second derivatives in the field equations.}\cite{pforms,bigal,multigal,solitons}  \be \label{mgal}
S_\text{multi-gal}= \int_{\cal M} d^D x  \sum_{m=1}^{D+1}  \alpha^{i_1 \dots i_m} \1 \del^{[a_2}\del_{a_2} \2 \cdots \del^{a_m]}\del_{a_m} \m  
\ee
where $\alpha^{i_1 \dots i_m} $ is symmetric. Here  $i,j, k$  labels the scalar, and $a, b, c$ labels the spacetime indices.  For $N$ scalars, $i,j,k$ run from $1\ldots N$, and in $D$ dimensions $a,b,c$ run from $0 \ldots D-1$. Throughout this paper, antisymmetrization omits the usual factor of $1/n!$.

{\it Covariantization} of (\ref{mgal}) is achieved by first minimally coupling the scalars to gravity which generically introduces higher order field equations. To restore the system to second order we add curvature dependent counter terms and arrive at the following
\be \label{covmgal}
S_\text{cov-multi-gal}=\int_{\cal M} d^D x \sqrt{-g}~\sum_{m=1}^{D+1} \alpha^{i_1 \dots i_m} \pi_{i_1} \nabla_{ a_{2}}\nabla ^{[a_{2}} \pi_{i_{2}} \cdots  \nabla_{a_{m}}\nabla^{a_{m}]} \pi_{i_{m}} + \sum_{m=3}^{D+1}\sum_{n=1}^{\lfloor \frac{m-1}{2}\rfloor }C_n^{m}  
\ee
where
\be
C_n^{m}  =\left(-\frac{1}{4 }\right)^{n}  \frac{(m-1)!}{(m-2n-1)!(n!)^2} \alpha^{i_1 \dots i_m}  \pi_{i_1} X_{i_2 i_3} \cdots X_{i_{2n} i_{2n+1}} \nabla_{ a_{2n+2}}\nabla ^{[a_{2n+2}} \pi_{i_{2n+2}} \cdots  \nabla_{a_{m}}\nabla^{a_{m}} \pi_{i_{m}} R^{b_1 c_1}{}_{b_1 c_1} \cdots  R^{b_n c_n]}{}_{b_n c_n}
\ee
for $n>0$, and  $X_{ij}=\frac{1}{2} \nabla_a \pi_{i} \nabla^a \pi_j$.

The covariant mutli-galileon theory (\ref{covmgal}) describes a multiple scalar-tensor theory, with potentially interesting applications, ranging from multi-field galileon inflation \cite{gal-inflation} to covariant self-tuning scenarios (see section \ref{sec:discussion}). In the case of a single scalar field, our theory does not quite reduce to the covariant galileon theory presented in \cite{covgal} owing to the fact that  the flat-space Lagrangians differ by a total derivative and this affects the subsequent covariant completion. Of course, both versions of the covariant single galileon still correspond to a subset of Horndeski's theory. The derivation of our covariant multi-galileon theory is presented in section \ref{sec:deriv}, with some details postponed to the appendix. The appendix also includes the resulting field equations.

In section \ref{sec:gen} we begin to generalise this theory, with a view to deriving a multi-scalar version  of Horndeski. Using methods similar to those presented in \cite{dgsz}, we first introduce the following generalised multi-galileon theory\footnote{Notice that, after some integration by parts,  (\ref{mgal}) is the  subset of (\ref{mscal})  with the functions $A$ and $A^{k_1 \ldots k_m}$ linear in $\bar X_{ij}$.}
\be \label{mscal}
S_\text{multi-scalar}= \int_{\cal M} d^D x ~A(\bar X_{ij}, \pi_l)+  \sum_{m=1}^{D-1} A^{k_1 \dots k_m} (\bar X_{ij}, \pi_l)\del^{[a_1}\del_{a_1} \pi_{k_2} \cdots \del^{a_m]}\del_{a_m} \pi_{k_m}
\ee
where $\bar X_{ij}=\frac{1}{2} \del_a \pi_{i} \del^a \pi_j$, and conjecture that this is the most general multi-scalar theory defined on Minkowski space, preserving second order field equations, provided $\frac{\del A^{i_1 \dots i_m}}{\del \bar X_{kl}}$ is symmetric in {\it all} of its indices $i_1, \ldots i_m, k, l$ (a formal proof of this conjecture now appears in \cite{proof}). We can covariantise this theory in the way described earlier, thereby arriving at the following
\begin{multline} \label{mhorn}
S_\text{cov-multi-scalar}=\int_{\cal M} d^D x \sqrt{-g}~A(X_{ij}, \pi_l)+  A^k(X_{ij}, \pi_l) \Box \pi_k \\
+\sum_{m=2}^{D-1}   \frac{(-4)^{\bar n } \bar n ! (m-2\bar n)! }{m !} \left[\frac{\del^{\bar n}}{\del X_{k_1 k_2} \cdots \del X_{k_{2\bar n-1} k_{2\bar n}}} B_m^{k_{2\bar n+1} \dots k_{m} }(X_{ij}, \pi_l)\right] \nabla_{ a_{1}}\nabla ^{[a_{1}} \pi_{k_{1}} \cdots  \nabla_{a_{m}}\nabla^{a_{m}]} \pi_{k_{m}} +\sum_{m=2}^{D-1}  \sum_{n=1}^{\bar n}Q_n^{m}  , 
\end{multline}
where $\bar n=\lfloor \frac{m}{2}\rfloor$ and
\ba
Q^m_n &=& \frac{(-4)^{\bar n -n} \bar n ! (m-2\bar n)! }{n! (m-2 n )!}\left[\frac{\del^{\bar n-n}}{\del X_{k_1 k_2} \cdots \del X_{k_{2(\bar n-n)-1} k_{2(\bar n-n)}} } B_m^{k_{2(\bar n-n)+1} \dots k_{m-2 n}} (X_{ij}, \pi_l)\right] \nonumber
\\
&&\qquad\qquad \nabla_{ a_{1}}\nabla ^{[a_{1}} \pi_{k_{1}} \cdots  \nabla_{a_{m-2n}}\nabla^{a_{m-2n}} \pi_{k_{m-2n}} R^{b_1 c_1}{}_{b_1 c_1} \cdots  R^{b_n c_n]}{}_{b_n c_n} 
\ea
Note that it is convenient  to rewrite $A^{k_1 \dots k_m}= \frac{(-4)^{\bar n } \bar n ! (m-2\bar n)! }{m!}  \frac{\del^{\bar n}}{\del X_{k_1 k_2} \cdots \del X_{k_{2\bar n-1} k_{2\bar n}}} B_m^{k_{2\bar n+1} \dots k_{m} }$ for $m \geq 2$, and we remind the reader that this should be taken to be symmetric in all of its indices, as should its higher derivatives with respect to $X_{ij}$. This generalised theory of multiple scalars and a single tensor reduces to Horndeski's theory for the case of a single scalar, and represents the maximal extension of the most general flat space theory in curved space. We conjecture that this theory would also be the most general multi-scalar tensor theory giving equations of motion of derivative order upto two, however a proof of this is not known yet. Again, the potential applications of this theory are likely to be considerable, from multi-field inflation to a possible multi-field extension of the {\it Fab-Four} \cite{fab4}. These and other future directions are discussed in greater detail in section \ref{sec:discussion}.
\section{multi-Galileons and covariantization} \label{sec:deriv}
We begin with the action describing multiple galileon fields in Minkowski space \cite{solitons},
\be
S_\text{multi-gal}= \int_{\cal M} d^D x  \sum_{m=1}^{D+1}  \alpha^{i_1 \dots i_m} \1 \del^{[a_2}\del_{a_2} \2 \cdots \del^{a_m]}\del_{a_m} \m  
\ee
where $ \alpha^{i_1 \dots i_m}$ is completely symmetric. Recall that anitsymmetrization omits the usual factor of $1/n!$ and that the indices $i,j, k$ label the scalar field, while $a, b, c$ are spacetime indices.  The first step towards covariantizing this theory is to couple gravity minimally, promoted partial derivatives to covariant ones, $\del_a \to \n_a$, such that 
\ba\label {actmincov}
S_\text{multi-gal} \to \int_{\cal M} d^D x \sqrt{-g} ~\sum_{m=1}^{D+1}  \alpha^{i_1 i_2 \dots i_m} \1 \n^{[a_2}{}_{a_2} \2 \dots \n^{a_m]}{}_{a_m} \m
\ea
Here we use the notation $\n_a{}^b \equiv \nabla_a \nabla^b$ and repeated indices are summed over.  Indeed, let us summarize the notation we will adopt for the remainder of this paper in the following table.
\begin{table}[h]
\begin{tabular*}{0.75\textwidth}{@{\extracolsep{\fill}}  l  l  l  }
 \hline \hline
Notation & Description & Definition/Example \\
\hline
$i,j,k \dots $ & Internal indices of the field & $\pi_i, \pi_j$ etc., \,\,  $i,j,k \in \{1 \dots N\}$ \\ [0.75 ex]
$a,b,c \dots$ & Space-time indices & $\n^a$\,\, $a,b,c \in \{ 0 \dots D-1 \}$ \\  [0.75 ex]
$\n_{ab} , \n^a{}_b$ & Double covariant derivative & $\n_{ab}\equiv  \n_a \n_b, \, \n_a{}^b \equiv \n_a \n^b $\\ [0.75 ex]
$I_{2p}, J_q$ & Collective unordered internal index & $I_{2p}\equiv \{ r_1 \dots r_{2p} \},\,J_q\equiv\{ s_1 \dots s_q\}$\\[0.6 ex]
&&Ex: $ A_{I_2 J_3}B^{I_2}C^{ J_3} = A_{i_1 \dots i_5} B^{i_1 i_2} C^{i_3 i_4 i_5} $\\ [0.75 ex]
$\a, \b, \c \dots $ & Antisymmetrized space-time index & $ X^{\a \b } \times Y^{\hat c \hat d \hat e} \times Z^{\hat f \hat g } \equiv X^{[a b} Y^{c d e} Z^{f g]}$ \\ [1 ex]
\hline\hline
\end{tabular*}
\label{tab:notations}
\caption{Notations}
\end{table}
It is also convenient  define the following scalars for the sake of brevity,
\ba
E_{I_{2p}} &=& \left(\n_{a_1} \pi_{r_1} \n^{a_1} \pi_{r_2}\right ) \dots \left(\n_{a_{p}} \pi_{r_{2p-1}}  \n^{a_{p}} \pi_{r_{2p}}\right)\\\nonumber
F_{J_q} &=& \left(\n_{a_1}{}^{\hat a_1}\pi_{s_1}\right) \dots \left(\n_{a_q}{}^{\hat a_q} \pi_{s_q} \right)\\\nonumber
G_{r} &=& R^{\hat a_1 \hat b_1}{}{}_{a_1 b_1} \dots R^{\hat a_r \hat b_r}{}{}_{a_r b_r}
\ea
Here we take $E_{I_0}=F_{J_0}=G_{0} =1$ and $E_{I_{2p}} = F_{J_{q}}=G_r = 0$ when $p,q,r$ are negative.
According to our notations we can write the $m$th order Lagrangian term as 
\be
 C_0^{m} \equiv \alpha^{i_1 i_2 \dots i_m} \1 \n^{[a_2}{}_{a_2} \2 \dots \n^{a_m]}{}_{a_m} \m  = \alpha^{i_1J_{m-1}} \1 F_{J_{m-1}} 
\ee
Variation of this term induced by $\pi_k \to \pi_k + \delta \pi_k$ where $k$ is an arbitrary integer between $1$ and $N$ is given by,
\ba
\delta C_0^{m} = \alpha^{k J_{m-1}} F_{J_{m-1}} \delta \pi_k + (m-1) \alpha^{i_1kJ_{m-2}} \1  \n_a{}^{\hat a} \delta \pi_k F_{J_{m-2}}
\ea
 and after integrating by parts we get,
\ba \label{eq:1}
\delta C_{0}^{m} &=& \bigg \{\alpha^{k J_{m-1}} F_{J_{m-1}} + (m-1) \alpha^{i_1kJ_{m-2}} \n_a{}^{\hat a} \1  F_{J_{m-2}}  \nn
&& + (m-1)(m-2) \alpha^{i_1ki_3 J_{m-3}}\big ( 2\,\n_{a} \1 \n^{\hat a \hat b}\n_{b} \3 F_{J_{m-3}}  + \1 \n_a \n^{\hat a \hat b} \n_b \3 F_{J_{m-3}}  \bigg \} \delta \pi_k \nn
&=&\bigg \{ \alpha^{k J_{m-1}} F_{J_{m-1}} + (m-1) \alpha^{i_1kJ_{m-2}} \n_a{}^{\hat a} \1  F_{J_{m-2}} + (m-1)(m-2) \alpha^{i_1 k i_3 J_{m-3}} \big ( \n_a \1 \n^c \3 R^{\hat a \hat b }{}{}_{bc} F_{J_{m-3}} \nn
&& - \frac{1}{4} \1 \n^c \3 (\n_c R^{\hat a \hat b}{}{}_{ab})  F_{J_{m-3}} + \frac{1}{2} \1 R^{\hat a \hat b}{}{}_{bc} \n_a{}^c \3 F_{J_{m-3}} \big ) \bigg \} \delta \pi_k
\ea
Here we have used the Riemann and Bianchi identities in the second step. To remove the term containing third derivatives in the metric we add the following counter term to the action.
\ba
C_1^{m} = -\frac{1}{8} \alpha^{i_1 I_2 J_{m-3}} \1 E_{I_2} F_{J_{m-3}} G_1
\ea
Although the variation of $E_{I_2}$ would generate the correct term to cancel the higher derivative term in (\ref{eq:1}, a further higher order term would be generated through the variation of $F_{J_{m-3}}$. Thus it is clear that a finite number of counter terms should be added recursively at each order in $\pi$. We find  that the counter term needed at the $n^\text{th}$ step is given by (see appendix for details),
\ba \label{Cnm}
C_n^m = T_n^{i_1 I_{2n} J_{m-2n-1}} \1 E_{I_{2n}} F_{J_{m-2n-1}} G_n
\ea
where,
\be \label{Tn}
T_n^{i_1 \dots i_m} = \left(-\frac{1}{8}\right)^{n}  \frac{(m-1)!}{(m-2n-1)!(n!)^2} \alpha^{i_1 \dots i_m} \qquad m\geq 3
\ee
It turns out that these counter terms are also sufficient to remove higher derivative terms generated in the $g_{ab}$ equation of motion (see appendix). Thus a covariant generalization of multi-galileon theory, preserving second order field equations, is given by
\ba\label{eq:cov}
S_\text{cov-mutli-gal} = \int d^D x \sqrt{-g} ~\sum_{m=1}^{D+1}  \sum_{n=0}^{\lfloor \frac{m-1}{2}\rfloor } C_n^{m} 
\ea
Of course, this was already expressed using more familiar notation in equation (\ref{covmgal}). The corresponding field equations are given by equations (\ref{pieq}) and (\ref{geqn}) in the appendix. For a single scalar field this theory does not quite reduce to the one presented in \cite{covgal}, although it does still correspond to a subset of Honrdeski's theory \cite{horny, dgsz}. The reason for the slight discrepancy is that our starting point in Minkowski space differs from that in \cite{covgal} by a total derivative and this affects the details of the subsequent covariantisation. 
\section{Towards Multi-scalar Horndeski} \label{sec:gen}
Having derived the covariant multi-galileon theory, it is natural to ask if we can go a stage further and find a multi-scalar generalisation of Horndeski's panoptic theory \cite{horny}. Recall that Horndeski's theory was rediscovered by DGSZ \cite{dgsz} using the following method: find the most general theory of a scalar in Minkowski space, with second order field equations, and then covariantise the resulting theory.  Here we will conjecture the form of the most general multi-scalar theory in Minkowski space, with second order equations of motion, and covariantise the result in order to give a generalised multi-scalar tensor theory of gravity. We do not attempt to prove the generality of our theory here, and leave this question as a future project.

To arrive at our proposed multi-scalar theory in Minkowski we begin by performing an integration by parts on the multi-galileon action (\ref{mgal}), and some relabelling, to arrive at the following
\be \label{mgal1}
S_\text{multi-gal}= \int_{\cal M} d^D x  ~-\frac{1}{2} \alpha^i \pi_i -\alpha^{ij} \bar X_{ij}-\sum_{m=1}^{D-1} \frac{m+2}{2} \alpha^{i j k_1\dots k_m} \bar X_{i j} \del^{[a_1}\del_{a_1} \pi_{k_1}  \cdots \del^{a_m]}\del_{a_m} \pi_{k_m}  
\ee
where we recall that $\bar X_{ij}=\frac{1}{2} \del_a \pi_{i} \del^a \pi_j$. An obvious generalisation of this, consistent with the one for a single scalar presented in \cite{dgsz}, is given by
\be \label{genmgal}
S_\text{multi-scalar}= \int_{\cal M} d^D x ~A(\bar X_{ij}, \pi_l)+  \sum_{m=1}^{D-1} A^{k_1 \dots k_m} (\bar X_{ij}, \pi_l)\del^{[a_1}\del_{a_1} \pi_{k_2} \cdots \del^{a_m]}\del_{a_m} \pi_{k_m}
\ee
Taken at face value, this action will yield higher order equations of motion. However, this can be avoided by imposing the condition that  $\frac{\del A^{i_1 \dots i_m}}{\del \bar X_{kl}}$ is symmetric in {\it all} of its indices $i_1, \ldots i_m, k, l$.  We conjecture that this theory is the most general multi-scalar theory in Minkowski space, with second order equations of motion. This is certainly true for the case of a single scalar for which (\ref{genmgal}) reduces to the general theory presented in \cite{dgsz}.

The next step is to covariantise the theory (\ref{genmgal}). As before we begin by minimally coupling to gravity, promoting partial derivatives to covariant ones,  $\del_a \to \n_a$ yielding 
\ba
S_\text{multi-scalar} \to \int d^Dx \sqrt{-g} \sum_{m=0}^{D-1} A(X_{ij}, \pi_l)^{i_1 \dots i_m} \n_{a_1}{}^{[a_1} \1  \dots \n_{a_m}{}^{a_m]} \pi_{i_m} 
\ea
with $X_{ij} := \frac{1}{2} \n_a\pi_{i} \n^{a} \pi_j$. Analogous to the covariantized multi-galileons, this action would yield equations of motion of derivative order greater than two. Thus we introduce the following counter terms at each order in $\pi_i$ to cancel those higher derivatives.
\ba
Q^m_n =A_n(X_{ij},\pi_l)^{J_{m-2n}} F_{J_{m-2n}} G_n 
\ea
Note that $Q_0^{m}=A(X_{ij}, \pi_l)^{i_1 \dots i_m} \n_{a_1}{}^{[a_1} \1  \dots \n_{a_m}{}^{a_m]} \pi_{i_m} $. In order to impose the constraint that the equations of motion are  second order, we take the variation of $Q^m_n$ induced by $\pi_k \to \pi_k + \delta \pi_k$. We focus only on those terms that contain higher derivatives, and using the following short-hand,
\ba
A_n^{J_{q}} = A_n(X_{ij},\pi_l)^{J_{q}}, \qquad \del^{ij} A_{n}^{J_q} = \frac{\del A_{n}^{J_q}}{ \del X_{ij}}
\ea
we obtain
\ba
\delta Q^m_n &=& \delta A_n^{J_{m-2n}} F_{J_{m-2n}} G_n + A_n^{J_{m-2n}} \delta F_{J_{m-2n}} G_n \nn
&=& \del^{ik} A_n^{J_{m-2n}} \n^a \delta \pi_k \n_a \pi_i F_{J_{m-2n}} G_n + (m-2n) A_n^{kJ_{m-2n-1}} \n_a{}^{\hat a} \pi_k F_{J_{m-2n-1}} G_n
\ea
After performing an integration by parts, and using both the Riemann and BIanchi identities, we find that $\delta Q^m_n$ contains the following terms that will contribute higher derivatives in the equations of motion,
\ba
\delta Q^m_n  &\supset& \bigg \{ - (m-2n) \del^{ik} A_n^{J_{m-2n-1} j} \n_a \pi_i \n^a \n^{\hat b} {}_{b} \pi_j F_{J_{m-2n-1}} G_n 
+(m-2n) \del^{ij} A_n^{J_{m-2n-1} k} \n_a \pi_i \n_b{}^{\hat b} \n^a \pi_i F_{J_{m-2n-1}} G_n \nn
&&-n\, \del^{ik} A_n^{J_{m-2n}} \n^a \pi_i (\n_a R^{\hat b \hat c}{}{}_{bc})  F_{J_{m-2n}} G_{n-1}  - \frac{(m-2n)(m-2n-1)}{4}\, A_n^{kiJ_{m-2n-2}}  \n^a \pi_i (\n_a R^{\hat b \hat c}{}{}_{bc}) F_{J_{m-2n-2}} G_n \bigg \} \delta \pi_k
\ea
It turns out that the first two terms cancel when we make use of the Riemann identity  $[\n_a \n_b]\n_c \pi_i = R_{abcd} \n^d \pi_i$. The cancellation follows  from the fact that we have the following constraint on the functions $A^{J_q}_n$,
\ba\label{eq:prop}
\del^{ij } A_n^{I_q } = \del^{(ij} A_{n}^{I_q )} \ea
where the bracket ($ij\dots$) stands for the symmetrization of the indices. We should also note that subsequent derivatives preserve this property of the function ie., $ \del^{ij} \del ^{kl} A_n^{J_q} = \del^{(ij} \del ^{kl} A_n^{J_q)}$. 

The remaining higher order terms are now
\ba\nonumber
\delta Q^m_n &\supset&  \bigg \{ -n\, \del^{ik} A_n^{J_{m-2n}} \n^a \pi_i (\n_a R^{\hat b \hat c}{}{}_{bc})  F_{J_{m-2n}} G_{n-1}  - \frac{(m-2n)(m-2n-1)}{4}\, A_n^{ikJ_{m-2n-2}}  \n^a \pi_i (\n_a R^{\hat b \hat c}{}{}_{bc}) F_{J_{m-2n-2}} G_n \bigg \} \delta \pi_k
\ea
These can be cancelled off by successive counter terms, $Q^m_{n+1}$, provided the following recursive relationship holds.
\ba\label{eq:recursive-gen}
(s+1) \del^{ij} A_{s+1}^{J_{m-2s-2}}  = -\frac{(m-2s)(m-2s-1)}{4} A_s^{ij J_{m-2s-2}} 
\ea
Operating with $(\del^{ij})^{s-n}$ on both sides and taking the following product, we find that 
\ba
\prod_{s=n}^{\bar n -1 } \left ( \frac{(\del^{kl})^{s-n} A_s^{ij J_{m-2s-2}}}{(\del^{kl})^{s-n} \del^{ij}A_{s+1}^{J_{m-2s-2}}} \right ) = \frac{A_n^{J_{m-2n}}}{(\del^{ij})^{\bar n-n}A_{\bar n }^{J_{m-2\bar n}}}=\frac{(-4)^{\bar n -n} \bar n ! (m-2\bar n)! }{n! (m-2 n )!}
\ea
Here $\bar n = \left \lfloor \frac{m}{2} \right \rfloor$  denotes the last counter term. We take $A_{\bar n}^{J_{m-2\bar N}} = B_{m}^{J_{m-2\bar n}}$ to define an arbitrary function for each $m$, giving,
\ba
A_n^{J_{m-2n}} = \frac{(-4)^{\bar n -n} \bar n ! (m-2\bar n)! }{n! (m-2 n )!}  (\del^{ij})^{\bar n - n} B_m^{J_{m-2\bar n}}           \quad  0 \leq n \leq \bar n
\ea
We conclude that the following generalized multi-scalar tensor theory is  has second order field equations from variation of the scalars
\ba
S_\text{cov-multi-scalar} = \int d^Dx \sqrt{-g} \sum_{m=0}^{D-1} \sum_{n=0}^{\left\lfloor \frac{m}{2} \right\rfloor} Q^m_n  
\ea
This action is written using more  familiar notation in equation (\ref{mhorn}).  It remains to show that it does not give rise to   higher derivatives in the $g_{ab}$ equations of motion. Variation of the metric gives
\ba
\delta Q^m_n  &\supset& A_n^{J_{m-2n}} \delta F_{J_{m-2n}} G_n + A_n^{J_{m-2n}} F_{J_{m-2n}} \delta G_n \nn
&=& (m-2n) A_n^{iJ_{m-2n-1}} \left( \n^{\hat a b} \pi_i \delta g_{ab} - \n_a \pi_i \delta g_{ab}{}{}^{;\hat a} + \frac{1}{2} g^{\hat a b} \n^c \pi_i \delta g_{ab;c} \right ) F_{J_{m-2n-1}}G_n \nn
&& + n \, A_n^{J_{m-2n}} F_{J_{m-2n}} \left ( R_c{}^{b\hat a \hat c} \delta g_{ab} - 2 g^{\hat a b} \delta g_{ab;c}{}{}{}^{\hat c} \right ) G_{n-1} 
\ea
where we again focus on terms that yield higher derivatives. Integrating by parts  and making use of the geometric identities, we find that
\ba
\delta Q^m_n &\supset& \bigg \{-\frac{(m-2n)(m-2n-1)}{2}\, A_n^{ijJ_{m-2n-2}} \n_c \pi_i \n^c \n^{\hat d}{}_d \pi_j F_{J_{m-2n-2}} G_n g^{\hat a b} \nn
&& -2n\,\del^{ij} A_n^{J_{m-2n}} \n_c \pi_i \n^c \n^{\hat d}{}_d \pi_jF_{J_{m-2n}} G_{n-1} g^{\hat a b} \bigg \} \delta g_{ab}
\ea
It is clear that these higher derivative terms would cancel if the same recursive relationship (\ref{eq:recursive-gen}) holds. We therefore conclude that our generalised theory (\ref{mhorn}) gives at most second order field equations under variation of all fields. As a consistency check of our work, it is reassuring to see that (\ref{mhorn}) does indeed reduce to Horndeski's theory \cite{horny,dgsz} in the case of a single scalar.

\section{Discussion} \label{sec:discussion}
In this paper, we have shown how gravity may be coupled to multi-galileons \cite{bigal,multigal,solitons}  without introducing higher order field equations,  and generalised our result, proposing a multi-scalar version of Horndeski's panoptic scalar-tensor theory \cite{horny,dgsz}. The actions for these theories are given by equations (\ref{covmgal}) and (\ref{mhorn}) respectively, and both may have interesting applications in multi-field inflation and quintessence scenarios. Indeed, the renaissance of Horndeski's single scalar-tensor theory has prompted a scan of generalised single field inflation models \cite{Kob}. Already there we see a variety of unexpected observational signatures eg. galileon inflation can give rise to observable 4-point functions even when the 3-point function is small \cite{effinfl}. The multi-scalar tensor theory proposed here now opens up the possibility of scanning the properties of generalised multi-field inflation models.

For the case of two galileons, the covariant theory (\ref{covmgal}) may be particularly relevant in the context of the cosmological constant problem. In \cite{bigal} it was shown that certain classes of bigalileon theories can give rise to self-tuning, with the physical spacetime screened from the cosmological constant b y the self-adjusting galileon fields. This is not in contradiction to Weinberg's famous no-go theorem \cite{nogo} as Poincar\'e invariance is broken in the galileon sector.  However, the proposal presented in \cite{bigal} was not entirely satisfactory for the following reason.  Self-tuning a large cosmological constant represents a considerable deviation from General Relativity. Whilst this is desirable on cosmological scales, one requires some mechanism through which deviations are screened in the solar system. In the bigalileon theory, this was achieved through the Vainshtein mechanism (see eg. \cite{vainsh}). However, if we want to self-tune a large cosmological constant and at the same time exploit Vainshtein screening, one finds that the decoupled bigalileon description breaks down. This begs the question, what happens in a covariant completion of these theories when one is no longer forced to work in the decoupling limit? The covariant multi-galileon theory will now enable us to address this issue and was one of the original motivations for this work.

We  will postpone a detailed analysis of these self-tuning scenarios. For now, let us present the covariant completion of the self-tuning bigalileon theory given as an example  in \cite{bigal}.  In the decoupling limit, this example is given by the action
\be
S=\int d^4 x ~\frac{M_{pl}^2}{2} \sqrt{-g} R +{\cal L}_{\pi, \xi} +S_m[e^{2\pi} g_{ab}; \Psi_n]
\ee
where the bigalileon Lagrangian is given by
\be
{\cal L}_{\pi, \xi}=3 M_{pl}^2 \pi \Box \pi-\frac{M^4}{\mu^2} \pi \Box \xi +\frac{1}{3 \mu^2} \pi \del_a\del^{[a} \pi \del_b{\del}^b \xi \del_c{\del}^{c]} \xi
\ee
and $S_m$ describes matter minimally coupled to the metric $e^{2\pi} g_{ab}$. The covariant completion of this theory can now be immediately read off from (\ref{covmgal}) and is given by
\begin{multline}
S=\int d^4 x ~ \sqrt{-g} \left[ \frac{M_{pl}^2}{2} R +3 M_{pl}^2 \pi \Box \pi-\frac{M^4}{\mu^2} \pi \Box \xi +\frac{1}{3 \mu^2} \pi \nabla_a\nabla^{[a} \pi \nabla_b{\nabla}^b \xi \nabla_c{\nabla}^{c]} \xi
 -\frac{1}{6\mu^2} \pi ( \n_a \pi \n^a \xi ) \n_b\n^{[b} \xi R^{cd]}{}{}_{cd}\right. \nn \left.
-\frac{1}{12 \mu^2} \pi ( \n_a \xi \n^a \xi)  \n_b \n^{[b} \pi R^{cd]}{}{}_{cd} \right]+S_m[e^{2\pi} g_{ab}; \Psi_n]
\end{multline}
It would be interesting to study the behaviour of this model  in some detail,  as well as the covariant completions of other self-tuning models presented in \cite{bigal}. How does self-tuning manifest itself? How large a cosmological constant can one tolerate and still be compatible with solar system tests?
  
Staying on the subject of self-tuning, we note that our proposed multi-scalar version of Horndeski's theory (\ref{mhorn}), puts us in a good  position to generalise the so-called {\it Fab Four}  theory \cite{fab4} to multiple fields. The {\it Fab Four} Lagrangians were obtained by asking which subset of Horndeski's theory can ``solve"  the cosmological constant problem in that they  screen the curvature from the vacuum energy. Given the generality of Horndeski, this enables one to  say that a self-tuning single scalar-tensor theory in four dimensions {\it must} correspond to a {\it Fab Four} theory. This is a rather powerful statement, but we are now in position to make it even more powerful by generalising to multiple fields. Mutliple fields will open up new possibilities as well, allowing for greater flexibility in deriving stable, phenomenologically consistent solutions. 

We close our discussion by drawing attention to an interesting two-scalar tensor theory which can now be seen as a subset of the two-scalar version of our generalised theory (\ref{mhorn}). This so-called {\it Fab Five} theory is an extension of  certain {\it Fab Four} Lagrangians\cite{fab5} and  is given by the following
\be
S_\text{Fab5}=\int d^4 x ~ \sqrt{-g} \left[ \frac{M_{pl}^2}{2} R-\frac{c_1}{2} (\nabla \pi)^2+f\left(-\frac{c_2}{2} (\nabla \pi)^2+\frac{c_G}{M^2} G^{ab} \nabla_a \pi \nabla_b \pi \right) \right]
\ee
When the function $f$ is the identity, this corresponds to a theory built from John and George from the {\it Fab Four}, along with a canonical kinetic term for the scalar. Generalising $f$ introduces an additional scalar degree of freedom, and the theory can be written as \cite{fab5}
\be
S_\text{Fab5}=\int d^4 x ~ \sqrt{-g} \left[ \frac{M_{pl}^2}{2} R-\frac{1}{2}(c_1+c_2 f'(\xi)) (\nabla \pi)^2+\frac{c_G}{M^2} f'(\xi) G^{ab}\nabla_a \pi \nabla_b \pi +f(\xi)-\xi f'(\xi)\right]
\ee
This corresponds to a particular two scalar-tensor theory  contained within (\ref{mhorn}). 

Whilst we have alluded to a few, at this stage it is impossible to envisage all the potential applications of our generalised mutli-scalar tensor theory. Horndeski's theory is currently the focus of plenty of research,  and to that we can now add its generalisation. See \cite{saltas} for some interesting recent use of Horndeski's theory that  may now be extended to include the theory presented in this paper.
\acknowledgements{AP was funded by a Royal Society University Research Fellowship and VS by a University of Nottingham International Office Fellowship.}

\appendix
\section{Recursive cancellation of higher order terms via counter terms}
Here demonstrate how the counter-terms defined for the covariant multi-galileon theorem, $C^m_n$ give rise to recursive cancellation of higher order derivatives upon variation. The methods described here were also applied to the generalised Horndeski theory, although the details are slightly different. Note that in passing we will present the field equations for the covariant multi-galileon theory (\ref{covmgal}) 
\subsection{$\pi_k$ equation of motion}
Let us begin with the scalar equations of motion. The minimally coupled Lagrangian term at $m^{th}$ order in $\pi_r$ is given by, 
\ba
C_{0}^m= \alpha^{i_1 \dots i_m} \1 \n_{a_2}{}^{[a_2} \2 \dots \n_{a_m}{}^{a_m]} \m
\ea
More generally we define  the corresponding counter term required at the $n^{th}$ recursive step at teh same order to be ,
\ba
C_n^m = T_n^{i_1 I_{2n} J_{m-2n-1}} \1 E_{I_{2n}} F_{J_{m-2n-1}} G_n
\ea
where $T_n^{i_1 I_{2n} J_{m-2n-1}}$ is symmetric in the last $(m-1)$ indices and to be determined.  Variation of $C_n^m$ induced by the variation in $\pi_k$ is,
\ba
\delta C_n^m &=&  T_n^{k I_{2n} J_{m-2n-1}} E_{I_{2n}} F_{J_{m-2n-1}} \delta \pi_k + 2n \,T_n^{i_1 k i_3 I_{2n-2} J_{m-2n-1}} \1 \n_a \3 E_{I_{2n-2}} F_{J_{m-2n-1}} G_{n} \n^a \delta \pi_k \nn
&&+  (m-2n-1) T_n^{i_1 k I_{2n} J_{m-2n-2}} \1 E_{I_{2n}} F_{J_{m-2n-2}} G_{n} \n_a{}^{\hat a} \delta \pi_k  
\ea
and subsequent integration by parts yields,
\ba
\delta C_n^m &=& \bigg \{ T_n^{k I_{2n} J_{m-2n-1}} E_{I_{2n}} F_{J_{m-2n-1}} G_n \nn
&& - 2n\, T_n^{i_1ki_3 I_{2n-2} J_{m-2n-1}} \n_a(\1 \n^a \3 E_{I_{2n-2}}) F_{J_{m-2n-1}} G_n  \nn
&& +2n(m-2n-1)\, T_n^{i_1 k i_3 i_4 I_{2n-2} J_{m-2n-2}} \bigg [-\1 \n_b \3 R^{b \a}{}{}_{ac} \n^c \4 E_{I_{2n-2}} F_{J_{m-2n-2}} G_n \nn
&&+ 2 \n_a \1 \n^{b \a} \3 \n_b \4 E_{I_{2n-2}} F_{J_{m-2n-2}} G_n  + \1 (\n_{ab} \3 \n^{\a b} \4 ) E_{I_{2n-2}} F_{J_{m-2n-2}} G_n \bigg ]\nn
&& + T_n^{i_1 k I_{2n} J_{m-2n-2}} (m-2n-1) \n_a{}^{\a} \1 E_{I_{2n}} F_{J_{m-2n-2}} G_n \nn
&& +2 \,{n \choose 2} (m-2n-1) T_n^{i_1 k i_3 i_4 i_5 i_6 I_{2n-4} J_{m-2n-2}} \1 (\n_{ab} \3 \n^b \4 \n_{c}{}^{\a} \5 \n^c \6 ) E_{I_{2n-4}} F_{J_{m-2n-2}} G_n \nn
&& \frac{m-2n-1}{2} {m-2n-2 \choose 2} T_n^{i_1 k i_3 i_4 I_{2n} J_{m-2n-4}} \1 (R^{\b \a}{}{}_{ad} \n^d \3 ) ( R_{bc}{}{}^{p(c)e} \n_e \4 ) E_{I_{2n}} F_{J_{m-2n-4}} G_n \nn
&&-2\,T_n^{i_1 k i_3 I_{2n} J_{m-2n-3}} \1 \n_a^c\3 R^{\a \b}{}{}_{bc} E_{I_{2n}} F_{J_{m-2n-3}} G_n \nn
&&- \frac{(m-2n-1)(m-2n-2)}{4}\,T_n^{i_1k i_3 I_{2n} J_{m-2n-3}} (\1 \n^c \3 \n_cR^{\a \b}{}{}_{ab}) E_{I_{2n}} F_{J_{m-2n-3}} G_n \nn
&&-2n^2\,T_n^{i_1 k i_3 I_{2n-2} J_{m-2n-1}} \1 \n_c \3 (\n^c R^{\a \b}{}{}_{ab}) E_{I_{2n-2}} F_{J_{m-2n-1}} G_{n-1} \bigg \} \, \delta \pi_k
\ea
Notice that the last two terms contain third derivative terms in the metric,
\ba
\delta C_n^m &\supset& - \bigg \{ \frac{(m-2n-1)(m-2n-2)}{4}\,T_n^{i_1k i_3 I_{2n} J_{m-2n-3}} (\1 \n^c \3 \n_cR^{\a\b}{}{}_{ab}) E_{I_{2n}} F_{J_{m-2n-3}} G_n \nn
&& -2n^2\,T_n^{i_1 k i_3 I_{2n-2} J_{m-2n-1}} \1 \n_c \3 (\n^c R^{\a \b}{}{}_{ab}) \3 E_{I_{2n-2}} F_{J_{m-2n-1}} G_{n-1} \bigg \} \, \delta \pi_k 
\ea
It is clear that these terms can be absorbed into each other recursively if the following relationship holds,
\ba \label{recursive}
\frac{T_{s+1}^{i_1 \dots i_m}}{ T_s ^{i_1 \dots i_m}} = -\frac{1}{8}\frac{(m-2s-1)(m-2s-2)}{(s+1)^2}
\ea
which implies that 
\ba
\prod_{s=0}^{n-1} \left [ \frac{T_{s+1}^{i_1 \dots i_m}}{ T_s ^{i_1 \dots i_m}} \right] = \frac{T_n^{i_1 \dots i_m}}{ \alpha^{i_1 \dots i_m}} = \left(-\frac{1}{8}\right)^n \frac{(m-1)!}{(n!)^2 \,(m-2n-1)!} \qquad m>2
\ea
This yields the result given by equation (\ref{Tn}). Finally, to express the $\pi_k$ equation of motion, we collect terms that at most second order  in $\delta C_n^m$. These are given by 
\ba
\epsilon_n^{(k)\,\,m} &=&  T_n^{k I_{2n} J_{m-2n-1}} E_{I_{2n}} F_{J_{m-2n-1}} G_n \nn
&& - 2n\, T_n^{i_1ki_3 I_{2n-2} J_{m-2n-1}} \n_a(\1 \n^a \3 E_{I_{2n-2}}) F_{J_{m-2n-1}} G_n  \nn
&& +2n(m-2n-1)\, T_n^{i_1 k i_3 i_4 I_{2n-2} J_{m-2n-2}} \bigg [-\1 \n_b \3 R^{b \a}{}{}_{ac} \n^c \4 E_{I_{2n-2}} F_{J_{m-2n-2}} G_n \nn
&&+ 2 \n_a \1 \n^{b \a} \3 \n_b \4 E_{I_{2n-2}} F_{J_{m-2n-2}} G_n  + \1 (\n_{ab} \3 \n^{\a b} \4 ) E_{I_{2n-2}} F_{J_{m-2n-2}} G_n \bigg ]\nn
&& +T_n^{i_1 k I_{2n} J_{m-2n-2}} (m-2n-1) \n_a{}^{\a} \1 E_{I_{2n}} F_{J_{m-2n-2}} G_n \nn
&&+ 2\,{n \choose 2} (m-2n-1) T_n^{i_1 k i_3 i_4 i_5 i_6 I_{2n-4} J_{m-2n-2}} \1 (\n_{ab} \3 \n^b \4 \n^{ \a }{}_{c} \5 \n^c \6 ) E_{I_{2n-4}} F_{J_{m-2n-2}} G_n \nn
&& \frac{m-2n-1}{2} {m-2n-2 \choose 2} T_n^{i_1 k i_3 i_4 I_{2n} J_{m-2n-4}} \1 (R^{\b \a}{}{}_{ad} \n^d \3 ) ( R_{bc}{}{}^{p(c)e} \n_e \4 ) E_{I_{2n}} F_{J_{m-2n-4}} G_n \nn
&&-2\,T_n^{i_1 k i_3 I_{2n} J_{m-2n-3}} \1 \n_a{}^c\3 R^{\a \b}{}{}_{bc} E_{I_{2n}} F_{J_{m-2n-3}} G_n \nn
\ea
It follows that the  $\pi_k$ equation of motion is given by,
\ba \label{pieq}
\sum_{m=1}^{D+1} \sum_{n=0}^{\lfloor \frac{m-1}{2} \rfloor} \epsilon_n^{(k)\,\,m} =0 
\ea
which is, of course, at most second order in derivatives, as desired.

\subsection{$g_{ab}$ equation of motion}
We now verify that our chosen counter-terms (\ref{Cnm}) also guarantee second order equations of motion from metric variation. To this end, we first note the following identities
\ba
\delta R^{\a \b}{}{}_{ab} X^{\hat c \hat d \cdots} &=&( \delta g^{\b c} R^{\a}{}_{cab} - 2 g^{\b c} \delta g_{bc;a}{}{}{}^{\a}) X^{\hat c \hat d \cdots} \\
\delta \n_a{}^{\hat a}\pi_{s} &=& - \n^{b\a} \pi_s \delta g_{ab} - \n^a \pi_s \delta g_{ab}{}^{;\a} + \frac{1}{2} g^{b \a} \n^c \pi_s \delta g_{ab;c}
\ea
The variation of the counter-term induced by the metric variation is,
\ba
\delta C_n^m &=& T_n^{i_1 I_{2n} J_{m-2n-1}} \left ( \1 \delta E_{I_{2n}} F_{J_{m-2n-1}} G_n + \1 E_{I_{2n}} \delta F_{J_{m-2n-1}} G_n + \1 E_{I_{2n}}F_{J_{m-2n-1}} \delta G_n \right ) \nn
&=& -T_n^{i_1 i_2 i_3 I_{2n-2} J_{m-2n-2}} \1 \n^a \2 \n^b \3 E_{I_{2n-2}} F_{J_{m-2n-1}} G_n \delta g_{ab} \nn
&&+ (m-2n-1) \, T_n^{i_1 i_2 I_{2n} J_{m-2n-2}} \1 E_{I_{2n}} F_{J_{m-2n-2}} G_n \big ( -\n^{b\a} \2 \delta g_{ab} - \n^b \delta g_{ab}{}{}^{;\a} + \frac{1}{2} g^{\a b} \n^c \2 \delta g_{ab;c} \big) \nn
&& +n\, T_n^{i_1 I_{2n} J_{m-2n-1}} \1 E_{I_{2n}} F_{J_{m-2n-1}} G_{n-1} \left ( R_c{}^{b \a \c} \delta g_{ab} - 2 g^{\a b} \delta g_{ab;c}{}{}{}^{\c} \right)
\ea
so that after integration by parts we obtain,
\ba
\delta C_n^m &=& \bigg \{T_n^{i_1 i_2 i_3 I_{2n-2} J_{m-2n-1}} \big [ -n \1 \n^a \2 \n^b  \3 E_{I_{2n-2}} F_{J_{m-2n-1}} G_n - 4n^2\, \1 \n_{cd} \2 \n^{\c d } \3 E_{I_{2n-2}} F_{J_{m-2n-1}} G_{n-1} g^{\a b} \nn
&& -4n^2 \, \1 \n^{\c} \n_{cd} \2 \n^d \3 E_{I_{2n-2}} F_{J_{m-2n-1}} G_{n-1} g^{\hat a b} \big ] \nn
&& +(m-2n-1) T_n^{i_1 i_2 I_{2n} J_{m-2n-2}} \big [ - \1 \n^{b \a} \2 E_{I_{2n}} F_{J_{m-2n-2}} G_n + (\1 \n^b \2 E_{I_{2n}})^{;\a} F_{J_{m-2n-2}} G_n \nn
&&- \frac{1}{2} (\1 \n^c \2 E_{I_{2n}})_{;c} F_{J_{m-2n-2}} G_n g^{\a b} - 2n\,  R^{ \c \hat d }{}{}_{d e} \n^e \2 \n_c ( E_{I_{2n}} \1 ) F_{J_{m-2n-2}} G_{n-1} g^{\a b} \nn
&&-n\, \1 \n_d{}^c \2 R^{\hat d \hat e} {}{}_{ec} E_{I_{2n}} F_{J_{m-2n-2}} G_{n-1} g^{\a b} \big ] \nn
&& +T_n^{i_1 i_2 i_3 I_{2n} J_{m-2n-3}} \bigg[ \frac{(m-2n-1)(m-2n-2)}{2} \1 \n^b \2 R^{\a \c}{}{}_{c d} \n^d \3 E_{I_{2n}} F_{J_{m-2n-3}} G_n \nn
&& - \frac{(m-2n-1)(m-2n-2)}{2} \1 \n^c \2 R_{cd}{}{}^{\hat d e} \n_e\3 E_{I_{2n}} F_{J_{m-2n-3}} G_n g^{\a b} \nn
&& - n {m-2n-1 \choose 2} \1 R^{\hat c \hat d}{}{}_{de} \n^e \2 R_{cf}{}{}^{ \hat f g} \n_g \3 E_{I_{2n}} F_{J_{m-2n-3}} G_{n-1} g^{\a b} \bigg] \nn
&&+ T_n^{i_1 I_{2n} J_{m-2n-1}} \big [ n \1 E_{I_{2n}} F_{J_{m-2n-1}} G_{n-1} R_c{}^{b \a \c } - 2n \n_c^{\c} \1 E_{I_{2n}} F_{J_{m-2n-1}} G_{n-1} g^{\a b} \nn
&& - 4n\, \n_c \1 \n^c E_{I_{2n}} F_{J_{m-2n-1}} G_{n-1} g^{\hat a b}\big]\nn
&&-T_n^{i_1 i_2 i_3 I_{2n} J_{m-2n-3}} \frac{(m-2n-1)(m-2n-2)}{2} \1 \n^{\c} \n_{cd} \2 \n^d \3 E_{I_{2n}} F_{J_{m-2n-3}} G_n g^{\a b}  \nn
&& -4n {n \choose 2} T_n^{i_1 i_2 i_3 i_4 i_5 I_{2n-4} J_{m-2n-1}} \1 \n^{\hat c}(\n_d \2 \n^d \3 ) \n_c ( \n_e \4 \n^e \5 ) E_{I_{2n-4}} F_{J_{m-2n-1}} G_{n-1} g^{\hat a b}   \bigg \} \delta g_{ab}
\ea
Again, focussing on the third derivative terms,
\ba
\delta C_n^m &\supset& \bigg \{-4n^2 T_n^{i_1 i_2 i_3 I_{2n-2} J_{m-2n-1}} \1 \n^{\c} \n_{cd} \2 \n^d \3 E_{I_{2n-2}} F_{J_{m-2n-1}} G_{n-1} g^{p(a)b} \nn
&& -T_n^{i_1 i_2 i_3 I_{2n} J_{m-2n-3}} \frac{(m-2n-1)(m-2n-2)}{2} \1 \n^{\c} \n_{cd} \2 \n^d \3 E_{I_{2n}} F_{J_{m-2n-3}} G_n g^{\a b}  \bigg \} \delta g_{ab}
\ea
we see that they can  are recursively cancelled if the same  relationship (\ref{recursive}) holds. As before, to express the metric equation of motion we collect terms up to second order in derivatives, remembering to include  the term generated by the variation of the metric determinant $\sqrt{-g}$. We find that the $g_{ab}$ equations of motion are given by,
\ba \label{geqn}
\sum_{m=1}^{D+1} \sum_{n=0}^{\lfloor \frac{m-1}{2} \rfloor} {\cal E}_n^{(m) ab} =0
\ea
where $
{\cal E}_n^{(m) ab} = \frac{1}{2} \left(  \epsilon_{n}^{(m) ab} + \epsilon_{n}^{(m) ba}      \right)
$
and
\ba
\epsilon_{n}^{(m) ab} &=& -n\, T_n^{i_1 i_2 i_3 I_{2n-2} J_{m-2n-1}} \1 \n^a \2 \n^b  \3 E_{I_{2n-2}} F_{J_{m-2n-1}} G_n - 4n^2\, \1 \n_{cd} \2 \n^{\c d } \3 E_{I_{2n-2}} F_{J_{m-2n-1}} G_{n-1} g^{\a b} \nn
&& +(m-2n-1) T_n^{i_1 i_2 I_{2n} J_{m-2n-2}} \bigg[ - \1 \n^{b \a} \2 E_{I_{2n}} F_{J_{m-2n-2}} G_n + (\1 \n^b \2 E_{I_{2n}})^{;\a} F_{J_{m-2n-2}} G_n \nn
&&- \frac{1}{2} (\1 \n^c \2 E_{I_{2n}})_{;c} F_{J_{m-2n-2}} G_n g^{\a b} - 2n\,  R^{ \c \hat d }{}{}_{d e} \n^e \2 \n_c ( E_{I_{2n}} \1 ) F_{J_{m-2n-2}} G_{n-1} g^{\a b} \nn
&&-n\, \1 \n_d^c \2 R^{\hat d \hat e} {}{}_{ec} E_{I_{2n}} F_{J_{m-2n-2}} G_{n-1} g^{\a b} \bigg] \nn
&& +T_n^{i_1 i_2 i_3 I_{2n} J_{m-2n-3}} \bigg[ \frac{(m-2n-1)(m-2n-2)}{2} \1 \n^b \2 R^{\a \c}{}{}_{c d} \n^d \3 E_{I_{2n}} F_{J_{m-2n-3}} G_n \nn
&& - \frac{(m-2n-1)(m-2n-2)}{2} \1 \n^c \2 R_{cd}{}{}^{\hat d e} \n_e\3 E_{I_{2n}} F_{J_{m-2n-3}} G_n g^{\a b} \nn
&& - n {m-2n-1 \choose 2} \1 R^{\hat c \hat d}{}{}_{de} \n^e \2 R_{cf}{}{}^{ \hat f g} \n_g \3 E_{I_{2n}} F_{J_{m-2n-3}} G_{n-1} g^{\a b} \bigg] \nn
&&+ T_n^{i_1 I_{2n} J_{m-2n-1}} \bigg[ n \1 E_{I_{2n}} F_{J_{m-2n-1}} G_{n-1} R_c{}^{b \a \c } - 2n \n_c^{\c} \1 E_{I_{2n}} F_{J_{m-2n-1}} G_{n-1} g^{\a b} \nn
&&- 4n\, \n_c \1 \n^c E_{I_{2n}} F_{J_{m-2n-1}} G_{n-1} g^{\hat a b}\bigg]  + \frac{1}{2} g^{ab} T_n^{i_1 I_{2n} J_{m-2n-1}} \1 E_{I_{2n}} F_{J_{m-2n-1}} G_n\nn
&&  -4n {n \choose 2} T_n^{i_1 i_2 i_3 i_4 i_5 I_{2n-4} J_{m-2n-1}} \1 \n^{\hat c} ( \n_d \2 \n^d \3 ) \n_c ( \n_e \4 \n^e \5 ) E_{I_{2n-4}} F_{J_{m-2n-1}} G_{n-1} g^{\hat a b}
\ea


\appendix

\end{document}